\documentstyle[preprint,aps]{revtex}
\begin{document}
\draft
\vskip 0.5cm
%\preprint{McGill/96-XX}
\title{Temperature determination from the lattice gas model}
\author{S. Das Gupta \ and \ J. Pan}
\address{Department of Physics, McGill University \\
3600 University St., Montr\'{e}al, PQ, H3A 2T8 Canada}
\author{M. B. Tsang}
\address{National Superconducting Cyclotron Laboratory and Department of 
		Physics and Astronomy	\\
Michigan State University, East Lansing, MI 48824, USA}

%\date{August 1996}
\maketitle

\begin{abstract}
Determination of temperature from experimental data has become important in
searches for critical phenomena in heavy ion collisions.  Widely used methods 
are ratios of isotopes (which rely on chemical and thermal equilibrium), 
population ratios of excited states etc.  Using the lattice gas model we 
propose a new observable: $n_{ch}/Z$ where $n_{ch}$
is the charge multiplicity and $Z$ is the charge of the fragmenting system.
We show that the reduced multiplicity is a good measure of the average
temperature of the fragmenting system.
\end{abstract}

\vskip 0.5cm

\newpage

The extraction of a temperature from observables in a fragmenting system is 
a much studied problem in heavy ion collisions.  It was originally
estimated from the slopes of inclusive spectra of charged particles. 
However, such calibrations are suspect because of collective flow and
other dynamical effects. Other procedures were proposed.  
One method is to extract temperatures
from populations of excited states of emitted fragments[1].  The practical
problem here is that of detector efficiency.  If chemical and
thermal equilibria are achieved, one may obtain temperature information
more easily from a double isotope ratio defined by[2]
\begin{eqnarray}
R={Y(A_i,Z_i)/Y(A_i+\Delta A,Z_i+\Delta Z) 
	\over Y(A_j,Z_j)/Y(A_j+\Delta A,Z_j+\Delta Z)}
\end{eqnarray}
where $Y(A_i,Z_i)$ is the total yield of the emitted fragment with mass and 
charge number $A_i,Z_i$.  One chooses $\Delta A$ and $\Delta Z$ to be the 
same for both the numerator and denominator to cancel out the effects of
proton and neutron chemical potentials.  This method was used in a recent
paper[3].  In that work it was found that as a function of $E^*/A$, the 
excitation energy per particle, the temperature was particularly flat around
T=5 MeV.  This implied a maximum in the specific heat which could be the 
signature of a phase transition.  This observation has sparked renewed
interest in the methods that were used to extract temperatures [4-7]

Careful investigations reveal that the deduced temperature may depend 
upon the particular
nuclei used in eq. (1). Thus ambiguities remain.  These ambiguities
arise from effects of sequential decay and other processes neglected
by the simple theoretical model of Ref. [2]. It is extremely difficult
to calculate such corrections from theory hence a phenomenogical approach
was taken in Ref. [4] to estimate corrections.  That analysis reduces much
of the ambiguity. 

In this communication we propose the reduced multiplicity $m=n_{ch}/Z$
as a thermometer, where $n_{ch}$ is the charge particle multiplicity
and $Z$ is the charge of the fragmenting system.  At zero temperature $m$ is
of order zero and at high temperature it approaches  1.  
If $m$ monotonically increases with $T$, one can attempt to extract 
$T$ from a measurement of $m$.  
There are several reasons why this needs be studied.

(1) The temperature extracted in this approach is in some sense an averaged 
temperature exploiting the properties of
all fragments rather than a selected few which are used as thermometers.

(2)  Calculations show that $m$ is at least nearly, if not totally, 
independent of the size of the fragmenting system.  Thus we can use it for 
a range of systems of different sizes. 

(3)  In other investigations a parametrization of $m$ in terms of $T$ may 
become useful for estimating critical exponents[5].

(4)  Success of the lattice gas model depends on the ability of predicting
correctly overall yields at freeze-out density as the excitation energy per
particle (which is related to temperature) changes.  Our past successes
in correlating with data[8-10] suggest that the proposed method 
will be a reasonable measure of temperature. 

%%%% This paragraph is replaced 
%(5)  The problem of subsequent decay is either small or absent in the model.
%This is because of the way the sizes of clusters are determined in the lattice
%gas model.  Unlike the Copenhagen model[11] where clusters appear at a 
%given temperature, we ascribe momenta to nucleons at a given temperature and
%then decide whether two nucleons with their velocities will bind or not.
%The formula we use to deduce cluster size reduces a truly many-body problem
%into independent sums of two body problems.  This is an approximation.
%There will be scenarios where our prescription for clusters will overestimate
%the size of a cluster.  But there also are scenarios where our formula, 
%for example, obtains two clusters, whereas the two clusters will actually 
%coalesce into one.  Our calculation for clusters was compared in Ref. [12] 
%with a many body molecular dynamics calculation and statistically negligible 
%difference was found.
%
%

%%%% This is the new paragraph.
(5)  An important question is whether the numbers of clusters which are formed 
in such models will change a great deal by sequential decay.  We like to point
out that the formation of clusters in the lattice gas model is quite different
from that in some other models, for example, the Copenhagen model[11].  In the
latter clusters are formed at a given temperature.  A cluster at a finite
temperature will have components of many excited states and many of these
excited states can emit particles.  In the lattice gas model this is done 
differently.  From a global temperature we first generate by Monte-Carlo
simulation momentum of each nucleon.  We then decide with these momenta
whether two nucleons will bind or not.  Thus our clusters do not have
a temperature; they instead have a well-defined excitation energy.  Within
rules of classical mechanics they will not decay.  The formula we
use to deduce cluster sizes reduces a truly many-body problem into independent
sums of two body problems.  This is an approximation.  There will be scenarios
where our prescription for clusters will overestimate the size of a cluster.
But there are also scenarios where our formula, for example, obtains two
clusters, whereas the two clusters will actually coalesce into one.  Lattice
gas calculation for clusters was compared with a many body molecular dynamics
calculation in Ref. [12] and statistically negligible difference was found.

One might argue that actual nuclei are quantum mechanical objects and classical
calculations should be followed by a ad-hoc correction.  In this work 
however we stay entirely within the classical rules followed in the model
and obtain the consequences.
%end of change

The lattice gas model has been described before[8,9].  In our previous 
calculations we had used the same value of interaction $\epsilon$
between neutron-proton, neutron-neutron and proton-proton.  An attractive
interaction was used.  We now use an attractive interaction $\epsilon_{np}$
between neutron-proton but the bond between neutron-neutron and proton-proton
should be either zero or slightly repulsive.  This eliminates unphysical
clusters like di-neutrons or di-protons and makes the model much more 
realistic.  In accordance with a force much used in molecular dynamics
calculation for nuclear collisions[13] we have used a slightly repulsive 
potential for neutron-neutron and proton-proton bond.  Thus $\epsilon_{np}=
\epsilon$ and $\epsilon_{nn}=\epsilon_{pp}=-\epsilon /5$.  The 
value of $\epsilon$ is negative: $\epsilon$ is the only energy scale in the
lattice gas model.  The value of $\epsilon$ can be chosen from binding energy
considerations.  In an infinite nuclear matter $\epsilon$ is about -5.33 MeV. 
In real finite nuclei because of Coulomb interaction and surface effects a 
lower value is expected to get the correct binding energy.  
As in the case of one type of bonds the thermal critical 
temperature appears at $T_c=-1.1275\epsilon$[12].  We can thus
either use $\epsilon$  or  $T_c$ as an energy unit.  We will use the latter.

Fig. 1 shows the dependence of temperature as a function of the reduced 
multiplicity $m=n_{ch}/Z$ for two systems with different mass number, A, 
and charge number Z : A=137, Z=57 and A=87, Z=37. In the calculation we assumed
a freeze-out density $\rho_f\approx 0.4\rho_0$  where $\rho_0$ is the
density of normal nuclear matter. As shown by the solid and dashed lines in the 
figure, the temperature is largely independent of the size of the 
fragmenting system.  

To explore the effect of freeze-out density on the temperature dependence
on $m$, two Monte Carlo calculations for the fragmenting system with 
A=137, Z=57 for $0.26\rho_0$ (solid lines) and $0.4\rho_0$ (dashed lines)
are plotted in Figure 2.
The two densities correspond to the freeze-out density of light 
fragmenting systems[10] and medium fragmenting systems[8,9]
Notice that $dm/dT$ goes to zero
at both $T=0$ and $T=\infty$ limit.  All densities give the same
high temperature limit.  The reason that the slope is zero at the low temperature
limit is that a certain amount of energy will have to be supplied to the 
fragmenting system before it will split into two or more parts.  This is a 
quantum effect of the lattice gas model due to the quantization of binding 
energy and lattice space.

To compare with experimental data, it would be desirable to
have a simple formula so that for a given value of $m$ one can readily deduce
the temperature given by this model.  Motivated by this we have tried
a simple prametrization; 
\begin{eqnarray}
T/T_c=a \ln{b \sqrt{m} \over 1-\sqrt{m}}
\end{eqnarray}
The results of such parametrization 
$a=1/7.5$, $b=110$; and $a=1/5.3$, $b=40$ for low and high
density respectively, are shown as the two dot-dashed lines in Fig. 2. 
The parametrization agrees with the Monte-Carlo simulations up to 
$m=0.8$.  This is well within the range of temperature that are of 
current interest. We will ignore the region beyond $m=0.8$ here.
Similarly we ignore the region of very low $m$ ($\sqrt{m} < 1/b$).

Recently, temperatures have been determined as a function of total charge
particle multiplicities ($n_{tot}$) for the Au+C system at 1 GeV/A incident 
energies[5]. The analysis also extracted the fragment source size (A) and 
the multiplicities ($n_{ch}$) attributed to the emitted fragment source.
Assuming the fragment source has the same neutron to proton ratio 
as the Au projectile, the following relations are obtained
from Ref. [5], 

\begin{eqnarray}
n_{ch} =0.3 n_{tot}
\end{eqnarray}
\begin{eqnarray}
Z=-0.68 n_{tot}+80.5
\end{eqnarray}

As the temperatures measured with the hydrogen and helium isotope ratios
were more reliable and suffered less systematic uncertainties and 
sequential decay correction effects, we have plotted these data in
Fig. 2 by assuming $T_c=8.3$ MeV. The critical temperature
$T_c$ is estimated to be about 4-6 MeV depending on the size of the
lattice system. 
In the present study, for simplicity, $T_c$ is treated as an adjustable 
parameter. The slope of the data agrees with the dashed line which is 
the prediction for breakup density of $0.4\rho_0$. 
Over the region of $n_{ch}/Z$=0.1 to 0.4, there is 
good agreement between data and calculations. At very low multiplicities, 
the data deviates from the calculations. 
Instead of dropping rapidly to zero as in the calculations, the
slope of the data remains nonzero. This could
arise from the inability to accurately deduce both the multiplicity and
charge of the fragment source for very peripheral collisions. 

In summary, we have shown that the reduced charged particle multiplicity
can be used to extract the average temperature of the fragmenting system.
The extracted empirical parametrizations for eq. (2) that relates
the temperature to the reduced charged particle multiplicity for the
Au+C system at 1 GeV/A is $a=1/5.3$, $b=40$ and $T_c=8.3$ MeV. 

The authors would like 
to thank Dr. Hauger for supplying us with digital data of Ref. [5].
This work is supported by the Natural Sciences and Engineering Research
Council of Canada, by the Fonds FCAR of the Qu\'ebec Government
and by the National Science Foundation under Grant No. PHY-95-28844.

\newpage
\section*{Figure captions}

\begin{description}
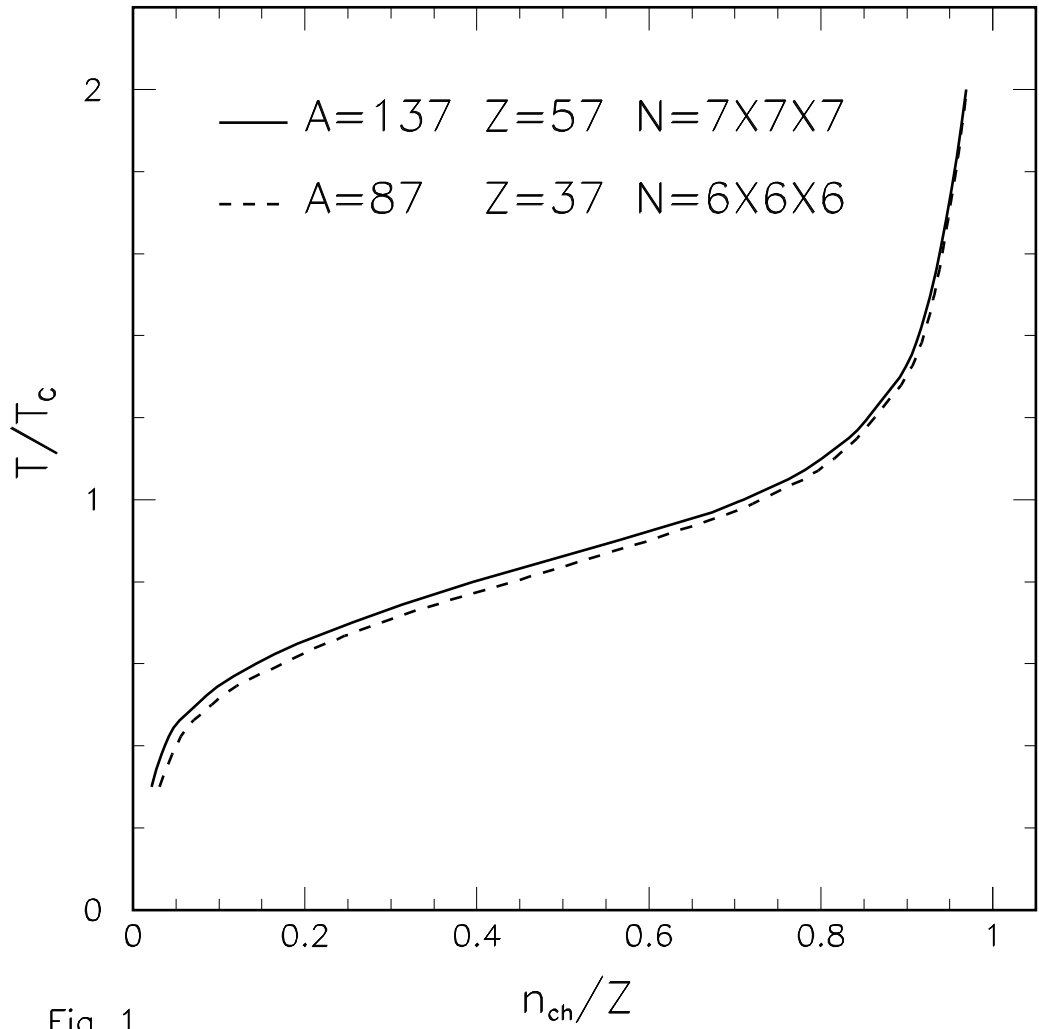
\item[Fig. 1]
Dependence of $T/T_c$ as a function of $n_{ch}/Z$ for fixed
freeze out density $\rho_f\approx 0.4\rho_0$ for two different fragmenting
systems.

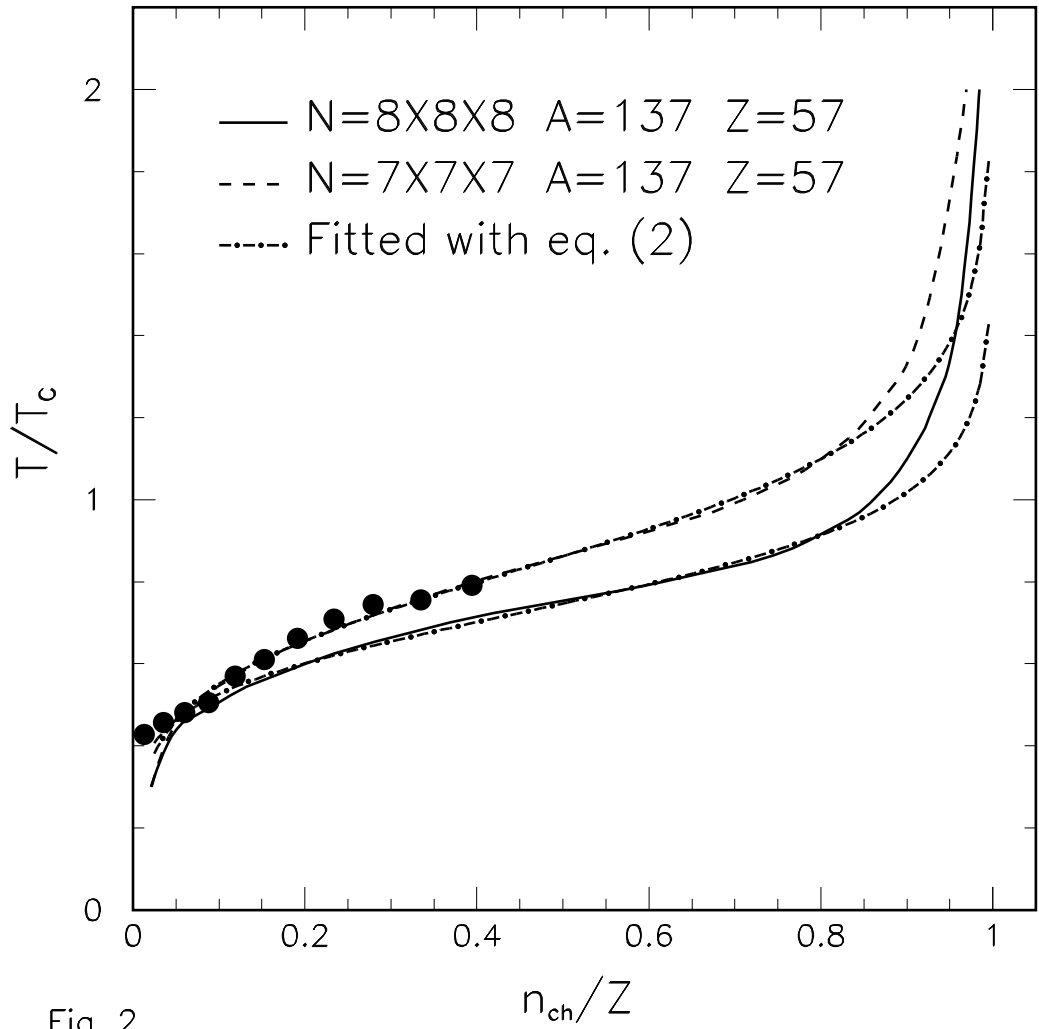
\item[Fig. 2] 
$T/T_c$ as a function of $n_{ch}/Z$ for two different 
freeze-out densities.  The solid and dashed lines are Monte Carlo
simulations with the Lattice gas models. The two dot dashed lines are
fits with eq. (2). Data points are the multiplicity data of Au+C 
reaction [5]. See text for detail.

\end{description}

\end{document}